\documentclass[aps,pre,twocolumn,10pt,superscriptaddress,nofootinbib,longbibliography,balancelastpage]{revtex4-1}
\usepackage[latin1]{inputenc}
\usepackage{amsmath,amssymb,amsfonts}
\usepackage{mathrsfs} 
\usepackage[capitalise]{cleveref}
\usepackage{siunitx}
\usepackage{braket}
\usepackage{calc}
\usepackage{longtable}%
\usepackage{tabularx}
\usepackage{dsfont}
\usepackage{color}
\usepackage{ifthen}
\usepackage[normalem]{ulem}% Fuer Durchstreichungen
\usepackage{tikz}
\usepackage{appendix}
\usetikzlibrary{shapes,arrows,positioning}

\allowdisplaybreaks

\newcommand{\Eins}{\mathds{1}}%
\newcommand{\ii}{\mathrm{i}}%
\newcommand{\HL}{H_{\mathrm{l}}}%
\newcommand{\HSL}{H_{\mathrm{sl}}}%
\newcommand{\dif}{\mathrm{d}}%
\newcommand{\tdif}[2]{\frac{\dif#1}{\dif#2}}%
\newcommand{\pdif}[2]{\frac{\partial#1}{\partial#2}}%
\newcommand{\Laplace}{\boldsymbol{\triangle}}%
\newcommand{\Tr}{\operatorname{Tr}}%
\newcommand{\ZT}[1]{\textquotedblleft#1\textquotedblright}%
\newcommand{\LS}{L_{\mathrm{S}}}%
\newcommand{\LH}{L_{\mathrm{H}}}%
\newcommand{\HS}{H_{\mathrm{S}}}%
\newcommand{\HH}{H_{\mathrm{H}}}%

\newcolumntype{L}[1]{>{\raggedright\arraybackslash}p{#1}}% 
\newcolumntype{C}[1]{>{\centering\arraybackslash}p{#1}}%
\newcolumntype{R}[1]{>{\raggedleft\arraybackslash}p{#1}}% 
\newcolumntype{Y}{>{\centering\arraybackslash}X}%    
\newcolumntype{Z}{>{\raggedright\arraybackslash}X}%  

\newlength{\myl}%
\newcommand{\SUM}[2]{{\setlength{\myl}{\widthof{$\displaystyle\sum_{#1}^{#2}$}*\real{0.5}-\widthof{$\displaystyle\sum$}*\real{0.5}}\sum_{#1}^{#2}\;\hspace{-\the\myl}}}% Summen in abgesetzten Gleichungen
\newcommand{\INT}[3]{\settowidth{\myl}{$\displaystyle\int_{#1}^{#2}$}{\int_{#1}^{#2}\;\;\;\hspace{-\the\myl}\dif #3}\,}% Integrale in abgesetzten Gleichungen
\newcommand{\TINT}[3]{\settowidth{\myl}{$\int_{#1}^{#2}$}{\int_{#1}^{#2}\!\ifthenelse{\equal{#1#2}{}}{}{\;\;\;\;\hspace{-\the\myl}}\dif #3}\,}%
\newcommand{\EINT}[3]{\settowidth{\myl}{$\int_{#1}^{#2}$}{\int_{#1}^{#2}\;\;\;\,\hspace{-\the\myl}\dif #3}\,}% Integrale in Exponenten

\begin{document}

\title{Projection operators in statistical mechanics: a pedagogical approach}
\author{Michael te Vrugt}
\affiliation{Institut f\"ur Theoretische Physik, Center for Soft Nanoscience, Westf\"alische Wilhelms-Universit\"at M\"unster, D-48149 M\"unster, Germany}

\author{Raphael Wittkowski}
\email[Corresponding author: ]{raphael.wittkowski@uni-muenster.de}
\affiliation{Institut f\"ur Theoretische Physik, Center for Soft Nanoscience, Westf\"alische Wilhelms-Universit\"at M\"unster, D-48149 M\"unster, Germany}

\begin{abstract}		
The Mori-Zwanzig projection operator formalism is one of the central tools of nonequilibrium statistical mechanics, allowing to derive macroscopic equations of motion from the microscopic dynamics through a systematic coarse-graining procedure. It is important as a method in physical research and gives many insights into the general structure of nonequilibrium transport equations and the general procedure of microscopic derivations. Therefore, it is a valuable ingredient of basic and advanced courses in statistical mechanics. However, accessible introductions to this method -- in particular in its more advanced forms -- are extremely rare. In this article, we give a simple and systematic introduction to the Mori-Zwanzig formalism, which allows students to understand the methodology in the form it is used in current research. This includes both basic and modern versions of the theory. Moreover, we relate the formalism to more general aspects of statistical mechanics and quantum mechanics. Thereby, we explain how this method can be incorporated into a lecture course on statistical mechanics as a way to give a general introduction to the study of nonequilibrium systems. Applications, in particular to spin relaxation and dynamical density functional theory, are also discussed.
\end{abstract}
\maketitle

\section{Introduction}
Although basic introductions to statistical mechanics tend to focus on thermodynamic equilibrium, a large part of modern research in this field is focused on nonequilibrium systems. These arise, e.g., in the study of driven and active\footnote{Active particles are particles that convert energy into directed motion. A good example are swimming microorganisms \cite{MarchettiJRLPRS2013}.} soft matter. Transport equations for systems of this type are obtained either from phenomenological considerations or from derivations starting from the underlying microscopic equations of motion (e.g., Hamilton's equations of motion). For working in this field, it is therefore important to have an understanding of both the structure of nonequilibrium transport equations and the way in which they can be derived.

It is a very remarkable and by no means obvious fact that the nonequilibrium dynamics of a very large class of systems can be described using only a small number of so-called \ZT{relevant} variables \cite{Uffink2006}. For example, fluids are already well characterized by mass, momentum, and energy density, although the number of microscopic degrees of freedom is much larger \cite{Forster1989}. Moreover, nonequilibrium transport equations all share a general form. They consist of a reversible and an irreversible part, where the irreversible part is, for close-to-equilibrium systems, proportional to the functional derivative of a free energy \cite{ThieleAP2012,Thiele2018,ThieleFHEKA2019}\footnote{This is not possible for active systems \cite{WittkowskiTSAMC2014}.}. Another point is that most of these transport equations do not contain memory terms. It is interesting to understand both why they usually do not arise and under which circumstances they do. Finally, sometimes these equations contain noise terms, whose origin from the underlying deterministic dynamics is also a conceptually interesting question.

An excellent way to study these aspects is the \textit{Mori-Zwanzig formalism} \cite{Nakajima1958,Zwanzig1960,Mori1965,Grabert1982,Zwanzig2001,teVrugtW2019}. It is a coarse-graining procedure, which allows to derive nonequilibrium transport equations in a systematic way from the underlying microscopic dynamics. The Mori-Zwanzig formalism has applications in a large number of fields, such as fluid mechanics \cite{Grabert1982, Forster1989}, dynamical density functional theory \cite{Yoshimori2005,EspanolL2009,AneroET2013, WittkowskiLB2012,WittkowskiLB2013, CamargodlTDZEDBC2018}, active matter physics \cite{LiluashviliOV2017}, spin relaxation theory \cite{KivelsonO1974, Bouchard2007,teVrugtW2019}, and particle physics \cite{Koide2002,HuangKKR2011}. 

The formalism, originally developed by Sadao Nakajima \cite{Nakajima1958}, Robert Zwanzig \cite{Zwanzig1960}, and Hajime Mori \cite{Mori1965}, exists in a large variety of forms and is known under different names, such as Zwanzig projection operator method \cite{Zeh1989}, Nakajima-Mori-Zwanzig formalism \cite{DominyV2017}, Mori-Zwanzig-Forster technique \cite{WittkowskiLB2012,WittkowskiLB2013}, or Kawasaki-Gunton method \cite{Yoshimori2005,KawasakiG1973}. Moreover, different forms differ in many technical details, although they all share the same general structure. Two main forms can be distinguished: methods where the projection operator is time-independent \cite{Mori1965,Zwanzig2001,Bouchard2007} and methods where it is time-dependent \cite{KawasakiG1973,Grabert1978,Grabert1982,MeyerVS2017,MeyerVS2019,teVrugtW2019}. 

This difference can form a major obstacle for students' understanding, since these methods are typically presented and derived in different ways. For time-independent methods, one introduces the projection operator through the idea of a Hilbert space formed by all observables, in which, through an appropriate scalar product, one can project onto a certain subspace formed by the relevant variables \cite{Zwanzig2001,Bouchard2007}. In the case of time-dependent projection operators, the idea of a Hilbert space or scalar product is used less frequently. Instead, the method is primarily discussed as a way to approximate the actual probability density of a system by means of a relevant probability density determined by macroscopically available information \cite{EspanolL2009}. Although both approaches are equivalent, and both approaches can be (and are) used in both cases, it requires a significant amount of calculation to see this (see Refs.\ \cite{Grabert1982,teVrugtW2019} for such calculations).

However, as we shall see, it is possible to turn this into an advantage. The simple form of the Mori-Zwanzig formalism, with time-independent projection operators, can be motivated and derived based solely on ideas that the students are familiar with from quantum mechanics. This gives important insights into many aspects of nonequilibrium statistical mechanics. Therefore, a simple derivation, as is presented in \cref{simple}, can already form an important part of an introductory course in statistical mechanics. Theories with time-dependent projection operators can then -- on their own or building up on this -- be explained in more advanced lectures. The relation between the different forms of the formalism is discussed in \cref{advanced}.

This article is structured as follows: In \cref{simple}, we derive the Mori-Zwanzig formalism with time-independent projection operators and discuss how this relates to general aspects of statistical mechanics. More advanced methods with time-dependent projection operators and Hamiltonians are discussed in \cref{advanced}. We summarize in \cref{conclusion}.

\section{\label{simple}Mori-Zwanzig formalism}
\subsection{General aspects of nonequilibrium transport equations}
In the description of a many-particle system in statistical mechanics, one is typically not interested in the precise coordinates of every single particle. Instead, a system is typically described using only a few relevant variables that follow closed dynamical equations of motion. A good example is an incompressible fluid, where, rather than calculating position and momentum of all fluid molecules, one describes the system using the flow field $\vec{u}(\vec{r},t)$ (or, equivalently, the momentum density $\varrho\vec{u}(\vec{r},t)$) with position $\vec{r}$ and time $t$. The flow field follows the famous Navier-Stokes equation
\begin{equation}
\pdif{\vec{u}}{t}+(\vec{u}\cdot\vec{\nabla})\vec{u} = - \frac{\vec{\nabla}p}{\varrho} + \nu \Laplace\vec{u},
\label{ns}%
\end{equation}
where $\varrho$ is the constant mass density, $p(\vec{r},t)$ the local pressure, and $\nu$ the kinematic viscosity of the fluid. 

Three things are notable here:
\begin{enumerate}
\item The complete system is described using only a single variable $\vec{u}(\vec{r},t)$. (This is based on the assumption of constant mass density and temperature, otherwise mass density and energy density would also come into play.) This variable is conserved, which implies that it varies on time scales that are much slower than those of microscopic fluctuations.
\item The equation consists of two parts. The first part, known as the \textit{Euler equation}, is essentially Newton's law for a fluid element (this is how it is often derived heuristically). It therefore describes reversible classical mechanics. The second part with the prefactor $\nu$ describes irreversible dissipative dynamics, which increases the entropy.
\item The equation is memory-less, i.e., the rate of change of $\vec{u}(\vec{r},t)$ depends only on the current value $\vec{u}(\vec{r},t)$, but not on values $\vec{u}(\vec{r},s)$ at previous times $s<t$. 
\end{enumerate}
Another example are the Bloch equations for a system of spins in a magnetic field $\vec{B}$. Using the total spin $\vec{S}$ as a relevant variable, it follows the Bloch equations
\begin{align}
\tdif{S_i}{t}= \gamma(\vec{S}\times\vec{B})_i - \frac{S_i}{\tau_i}    
\end{align}
with the gyromagnetic ratio $\gamma$ and the relaxation times $\tau_i$ with $i=x,y,z$. Again, we have a description that only requires one slow variable (although, this time, it is not conserved) and, again, we have a term for organized reversible motion (precession of the spins in the magnetic field) as well as a term for dissipation, describing the irreversible relaxation to equilibrium. Moreover, again, we have an equation that contains no memory effects.

\subsection{Mori theory}
We now wish to extend these observations to a general formalism that allows us to derive such equations from the underlying microscopic dynamics in a systematic way. This formalism is known as the \textit{Mori-Zwanzig formalism}. For introductory purposes, we present here only the simplest form (following Ref.\ \cite{Zwanzig2001}) and then discuss possible extensions. We present the method in quantum-mechanical form, since the classical case is essentially equivalent and not much simpler.

The main idea is that we can think of all the observables that we could use to describe the many-particle system as forming a Hilbert space. A Hilbert space, which should be familiar from quantum mechanics, is a vector space equipped with a scalar product.\footnote{Strictly speaking, this is a pre-Hilbert space. A Hilbert space additionally requires the convergence of every Cauchy sequence.} A simple example of a Hilbert space is $\mathbb{R}^{n}$ with the scalar product $(\vec{a},\vec{b}) = \vec{a}\cdot\vec{b}$. If we now have a vector $\vec{x}$ in a Hilbert space and are, for some reason, only interested in the part which points in the direction of a vector $\vec{c}$ (which does not have to be normalized), we can project $\vec{x}$ onto $\vec{c}$ by applying a projection operator $P$:
\begin{equation}
P\vec{x} = \frac{\vec{x}\cdot\vec{c}}{\vec{c}\cdot\vec{c}}\,\vec{c}.
\label{projection}%
\end{equation}

We now apply this idea to statistical mechanics. All our observables form a (very large-dimensional) Hilbert space. We are only interested in some particular observable $A$ (e.g., the momentum density in a fluid or the total magnetization in a spin system) that can be thought of as a \ZT{direction} in this large-dimensional Hilbert space. Defining an appropriate scalar product $(\cdot,\cdot)$, we then project an observable $X$ onto $A$ in the form
\begin{equation}
PX = (A,A)^{-1}(X,A)A.
\end{equation}
For a set of relevant variables $\{A_i\}$, this generalizes to\footnote{Throughout this article, we sum over each index appearing twice in a term.}
\begin{equation}
PX = (A_j,A_i)^{-1}(X,A_i)A_j.
\end{equation}
This form is constructed in complete analogy to \cref{projection}. (Note, however, that the form becomes more complicated for general projection operators, in particular if they are time-dependent.)

We are left with two tasks:
\begin{enumerate}
\item What is a good definition of a \ZT{scalar product} in the Hilbert space of dynamical variables?
\item How does this help us to describe the dynamics of the system?
\end{enumerate}

We start with the second task. For this, we need to think about what an observable is. In a classical system, it is a function $A(\vec{q},\vec{p})$ on the phase space with conjugate position $\vec{q}$ and momentum $\vec{p}$, while in quantum mechanics, it is a Hermitian operator acting on wave functions. In the Heisenberg picture, the observables are time-dependent while the wave functions are not. This is the picture we use here, since we are interested in the time-evolution of the observables. We assume, for simplicity, that our quantum observable $A$ is not explicitly time-dependent. Then, the Heisenberg equation of motion reads
\begin{equation}
\tdif{A}{t}= \frac{\ii}{\hbar}[H,A] = \ii L A,
\label{Heisenberg}%
\end{equation}
where we have introduced the imaginary unit $\ii$, the reduced Planck constant $\hbar$, the commutator $[\cdot,\cdot]$, the Hamiltonian $H$, and the corresponding Liouvillian $L$. In the classical case, we use the Poisson bracket $\{\cdot,\cdot\}$ instead of the commutator. The formal solution of \cref{Heisenberg} is the operator exponential
\begin{equation}
A(t) = e^{\ii Lt}A.
\end{equation}
To clarify the probably confusing notation, which is used also throughout the literature: By $A(t)$, we denote the time-dependent observable in the Heisenberg picture. This is related to the time-independent Schr\"odinger-picture observable by a time-dependent transformation (see Eq.\ \eqref{hs} further below). One has to choose a certain time (here we use $t=0$) at which the Heisenberg-picture observable coincides with the Schr\"odinger-picture observable \cite{BalianV1985}. By $A$ we denote the time-independent Schr\"odinger-picture observable. The same holds in the classical case, where a distinction between Heisenberg and Schr\"odinger picture also exists \cite{HolianE1985}, although it is less well known there. In the classical case, the Schr\"odinger-picture observable is a phase-space function $A(\vec{q},\vec{p})$, whereas the Heisenberg picture observable is a phase-space function $A(\vec{q}(t),\vec{p}(t))$. 

We now use the operator identity (\ZT{Dyson identity})
\begin{equation}
e^{\ii Lt} = e^{\ii QLt} + \INT{0}{t}{s}e^{\ii L(t-s)} P\ii Le^{\ii Q Ls}
\label{identity}%
\end{equation}
with the orthogonal projection operator $Q=1-P$,
which can be easily proven by differentiation (see \cref{dyson}). Applying \cref{identity} to $Q \ii LA_i $ gives the \textit{Mori-Zwanzig equation} 
\begin{equation}
\dot{A}_i(t) = \Omega_{ij} A_j(t) + \INT{0}{t}{s}K_{ij}(s)A_j(t-s) + F_i(t)
\label{exact}
\end{equation}
with the frequency matrix
\begin{equation}
\Omega_{ij} = (A_j,A_k)^{-1}(\ii L A_i,A_k),
\end{equation}
the memory matrix
\begin{equation}
K_{ij}(s)=(A_j,A_k)^{-1}(\ii L F_i(s),A_k),
\end{equation}
and the random force
\begin{equation}
F_{i}(t)=e^{\ii QLt} Q \ii LA_{i}.
\end{equation}
Equation \eqref{exact} only contains the relevant variables $\{A_k\}$, but it is still formally exact. We have achieved this by a memory term: The present rate of change $\dot{A}_i(t)$ depends not only on the present value $A_i(t)$, as described by the term containing $\Omega_{ij}$, but also on the values $A_i(s)$ at previous times $s<t$, which is described by the term containing $K_{ij}$. $F_i(t)$ is a noise term. It describes the influence of those variables that we are not interested in.

Since the exact result \eqref{exact} is a time-delayed integro-differential equation, it is, in general, extremely difficult to solve. We therefore make an approximation based on the assumption that the relevant variables are slow on macroscopic time scales. \ZT{Slow} means here that the rate of change is small. We can therefore make a Taylor expansion of \cref{exact} in $\ii L A_j$ up to second order. Since $K_{ij}$ is already of second order in $\ii L A_j$, we can write
\begin{equation}
A_j(t-s) = A_j(t) - \dot{A}_j(t)s + \dotsb
\end{equation}
and truncate the expansion after the first term. As this expression is integrated over in Eq.\ \eqref{exact}, we have to assume that the memory kernel $K_{ij}(s)$ vanishes very quickly on macroscopic time scales. In this case, the integrand in \cref{exact} vanishes for large times and there is no problem in extending the integral to $s=\infty$\footnote{Strictly speaking, the integral is extended to a time $\tau_c$, which is much longer than the relaxation time scale $\tau_R$, but smaller than the time scale of recurrence. \ZT{Recurrence} is the fact that a Hamiltonian system will, after a sufficiently long time, always return arbitrarily close to its initial state. Since recurrence occurs, for macroscopic systems, on time scales vastly longer than the age of the universe, this is no practical problem \cite{Zeh1989} and we can simply set the upper integration boundary to infinity.}. The resulting \textit{Markovian approximation} gives the approximate transport equation 
\begin{equation}
\dot{A}_i(t) = \Omega_{ij} A_j(t) + D_{ij} A_j(t)+ F_i(t)
\label{approx}%
\end{equation}
with the dissipative matrix\footnote{This matrix is also referred to as \ZT{diffusion matrix} \cite{teVrugtW2019}.}
\begin{equation}
D_{ij}= \INT{0}{\infty}{s}K_{ij}(s).
\end{equation}
As a further simplification, we can replace within $K_{ij}$ the term $e^{\ii QLt}$ by $e^{\ii Lt}$, 
since \cite{Zwanzig2001}
\begin{equation}
\ii PLX \propto (\ii L X, A) = - (X,\ii LA),
\end{equation}
such that $\ii PLX$ is of order $\ii LA$ and can be dropped.

Equation \eqref{approx}, which is local in time, is much easier to solve than Eq.\ \eqref{exact}. Physically, the main difference is that we have replaced the memory integral by the simple term $D_{ij} A_j(t)$. The fact that Eq.\ \eqref{approx} only depends on the present state explains the name \ZT{Markovian approximation}: A \textit{Markov process} is, in stochastics, a random process that has no memory.

The other task we had to address is the scalar product. For this, we first require the notion of a phase-space distribution function $\rho(\vec{q},\vec{p},t)$. In the Gibbsian framework of statistical mechanics\footnote{There is an alternative conceptual framework known as Boltzmannian statistical mechanics, which is not based on ensembles \cite{Frigg2008}. This framework is of historical and conceptual interest, but not frequently used in practical calculations.}, where a many-particle system is described using an \textit{ensemble}, which is a hypothetical set of infinitely many copies of the system with different initial conditions, this function specifies the probability that a system that is chosen at random from this ensemble is in the microstate specified by the phase-space coordinates $\vec{q}$ and $\vec{p}$ \cite{Frigg2008}. In the quantum-mechanical case, where the state of a system is specified by a wave function $\ket{\psi}$, the many-particle system is instead described by the statistical operator $\hat{\rho} = \sum_{i} p_i \ket{\psi}_i\bra{\psi}_i$, where $p_i$ is the probability that the system is in the quantum state $\ket{\psi}_i$.

A standard choice is to use a generalized correlation function. For close-to-equilibrium quantum systems, this is given by the Mori product \cite{Grabert1982}
\begin{equation}
(X,Y) = \INT{0}{1}{\alpha}\Tr(\hat{\rho}Xe^{-\alpha\beta H}Y^\dagger e^{\alpha \beta H})
\end{equation}
with the quantum-mechanical trace $\Tr$, the statistical operator $\hat{\rho}$,  the thermodynamic beta $\beta=1/(k_{B}T)$ with the Boltzmann constant $k_{B}$ and absolute temperature $T$, and the dagger $^\dagger$ denoting the Hermitian adjoint. More generally, the scalar product can be based on a so-called \ZT{relevant density} (see \cref{relevant}). The idea here is that we replace the unknown actual density operator of the system by one that only depends on the mean values of relevant variables, gives the correct values for them, and maximizes the informational entropy. This relevant density is then used to calculate mean values and correlation functions \cite{Grabert1982,HansenMD2009,Forster1989}.

\subsection{Mori-Zwanzig equation in a general context}
If we take a look at the equation of motion \eqref{approx}, we see that we have arrived at the general form we were interested in:
\begin{enumerate}
    \item The equations of motion depend only on the relevant variables $\{A_i\}$.
    \item They consist of a part $\Omega_{ij}A_j$ that describes reversible organized motion and of a dissipative contribution $D_{ij}A_j$\footnote{From the general structure \eqref{approx}, it is not obvious that $\Omega_{ij}A_j$ is reversible, while $D_{ij}A_j$ is not. To see this, considerations on the symmetries of the coefficients $\Omega_{ij}$ and $D_{ij}$ under time reversal are required, which can be found in Ref.\ \cite{Grabert1982}.}.
    \item They are memoryless, i.e., they only depend on the current state of the system.
\end{enumerate}
Thus, \cref{approx} is a paradigmatic case for a transport equation in statistical mechanics. In fact, all relevant transport equations can be derived from this formalism. Moreover, we now have found a way to understand where the general structure described above comes from and under which conditions it holds.

Up to \cref{exact}, our derivation was formally exact. If we describe a system not based on all microscopic degrees of freedom, but only with a reduced set of variables, the price we have to pay is that we need to know about the state of the system at previous times to fully determine its temporal evolution. There are, however, certain circumstances under which this is not necessary, namely if the relevant variables are varying very slowly compared to other degrees of freedom. In this case, an approximate equation can be derived that gives a closed dynamics for the relevant variables without memory effects.

This requires that the set of relevant variables, which the formalism itself cannot determine, contains all slow macroscopic variables. For example, in the description of fluids, a reasonable choice is to use mass density, momentum density, and energy density \cite{Grabert1982}. If we restrict ourselves to one variable only -- say, the mass density -- we make the strong assumption that all other variables relax on the time scale on which the chosen variable changes \cite{EspanolL2009}.  

The formalism does also allow to handle the case in which memory effects \textit{are} relevant, since the memoryless case is just a particular approximation of the more general transport equation \eqref{exact} in which memory effects are present. Thus, memory kernel can be calculated systematically based on the formalism \cite{GrabertTH1977,MeyerVS2017,MeyerPS2019,ChorinHK2002}.

It is interesting that the Mori-Zwanzig formalism allows to derive irreversible macroscopic equations of motion from the reversible microscopic laws. As shown in Ref.\ \cite{Grabert1982}, the dissipative terms lead to a monotonous increase of entropy and thus allow to prove a H-theorem, as long as the Markovian approximation holds (and only then). Thus, the assumption of Markovian behavior of the macroscopic variables is central to the emergence of macroscopic irreversibility. Moreover, as shown in Ref.\ \cite{Zeh1989}, one also requires an assumption about initial conditions, namely that the irrelevant part of the probability density (see below) vanishes in the distant past. If we had made the same assumption for the future, we could have proven the unphysical statement that the entropy always decreases.

For this reason, projection operator methods have also attracted the attention of philosophers of physics who are interested in where, given the reversible microscopic physics, irreversibility comes from. Philosophers study projection operators due to the general insights they provide into the origin of irreversible transport equations \cite{Wallace2013,Wallace2015}. A matter of debate here is, roughly speaking, whether the fact that irreversible laws arise from coarse-graining methods, which are often justified by the disinterest in microscopic details, lead to some undesirable amount of subjectivity. After all, the fact that entropy always increases seems to be a physical matter of fact which is not related to our ignorance of microscopic details of a statistical-mechanical system. A recent philosophical discussion of this problem based on the Mori-Zwanzig projection operator method can be found in Ref.\ \cite{Robertson2018}. Moreover, it is discussed what precisely justifies the assumption of Markovian behavior \cite{Uffink2006}. For a more detailed presentation of the debate in philosophy of physics, with an emphasis on how it relates to the one in statistical mechanics, see Ref.\ \cite{teVrugt2020}.

\section{\label{advanced}Advanced forms}
Although the theory based on \cref{exact,approx} is already a very powerful and, in principle, formally exact method, it has certain drawbacks. The most important one is that \cref{approx} is always a linear equation. As discussed in Ref.\ \cite{Zwanzig2001}, the variables $A$ and $A^2$, although they are obviously related, are different variables in a Hilbert space of dynamical functions. Therefore, if we project onto $A$, the dependence on $A^2$ belongs to the orthogonal dynamics contained in memory and noise terms. 

Nevertheless, most transport equations relevant for out-of-equilibrium systems, such as the Navier-Stokes equation \eqref{ns}, are \textit{nonlinear}. These nonlinearities are very important for the dynamics of these systems, e.g., when studying pattern formation. Hence, it is a significant advantage if we are able to derive nonlinear equations. This is possible in two ways. One option is to project onto a larger set of relevant variables that also includes nonlinear functions of the $\{A_i\}$. This is the basic idea behind the derivation of Fokker-Planck equations within the Mori-Zwanzig formalism \cite{Zwanzig2001,Grabert1982}. The other alternative, which we will present here, is to use time-dependent projection operators \cite{Grabert1982}.

Another limitation of the standard method is that it is restricted to time-independent Hamiltonians. This is not the most general case, since time-dependent external fields can have an important influence on the microscopic structure and macroscopic dynamics of a system. Therefore, time-dependent Hamiltonians, discussed further below in \cref{td}, have become an active field of research in the context of Mori-Zwanzig theory, both for time-independent \cite{Bouchard2007} and time-dependent \cite{teVrugtW2019,MeyerVS2019} projection operators.

\subsection{\label{relevant}Relevant probability density}
In the following, we focus on the time-dependent projection operator method presented in the classical textbook by Grabert \cite{Grabert1982}. It is a very general method applicable to nonlinear far-from-equilibrium dynamics, allows to describe also the dynamics of fluctuations \cite{Grabert1978}, and has recently been extended to incorporate time-dependent Hamiltonians \cite{teVrugtW2019}.

The time-dependent projection operator method is closely related to the usual methodology of statistical mechanics and extends it towards nonequilibrium systems. In statistical mechanics, the configuration of a many-particle system is described through a probability distribution $\rho$, which in classical mechanics is a function on phase space and in quantum mechanics a Hermitian operator (\ZT{density matrix}). For equilibrium systems, $\rho$ is constructed by optimizing a thermodynamic potential or the entropy. When choosing the microcanonical ensemble, the entropy, defined as $S = -k_B \Tr(\rho \ln(\rho))$, is maximized with the constraint that the total energy of the system is fixed (or, more precisely, that the total energy is between $E$ and $E+\Delta E$ with $\Delta E/E \ll 1$). 

In the information-theoretic approach to statistical mechanics, pioneered by E.\ T.\ Jaynes \cite{Jaynes1957a}, this method of maximizing the entropy is given an epistemic justification: The probability distribution is introduced as a way incorporating what we know (and what we do not know) about the microscopic configuration of a system. It is, as he argues, rational to assign to all microscopic configurations that are, as far as we know, possible, the same probability. This is formalized by demanding that the probability distribution is chosen in such a way that it maximizes the Shannon entropy 
\begin{equation}
\sigma = - \sum_{i}p_i\ln (p_i)
\end{equation}
with the probabilities $p_i$ of the possible microscopic configurations. 
For example, if we consider a many-particle system about which we know nothing but the fact that it has total energy $E$, particle number $N$, and volume $V$, then we can maximize the information entropy, i.e., a measure for how indifferent we are with respect to missing information, and arrive at the microcanonical distribution. Likewise, if our macroscopic information is that we have an \textit{average} energy $E$, we recover, by the same procedure, the canonical distribution with the temperature $T$ arising from the Lagrange multiplier fixing the average energy \cite{Jaynes1957a}.

While this method thus gives the desired results for the equilibrium case, it can also be applied to nonequilibrium systems \cite{Jaynes1957b}. Assume that we have, like above, a set of relevant variables $\{A_i\}$. Let their mean values be $\{a_i(t)\}$. These mean values are the macroscopic information we have about our system. Then, we choose our \textit{relevant probability density} as \cite{Grabert1978}
\begin{equation}
\bar{\rho}(t) = \frac{1}{Z(t)}e^{-\lambda_i(t)A_i}.
\label{rhobar}%
\end{equation}
with the normalization $Z(t)$ and the thermodynamic conjugates $\{\lambda_i(t)\}$. 
This density maximizes the informational entropy with respect to the constraint that our macroscopic information is given by the mean values $\{a_i(t)\}$ of the relevant variables \cite{AneroET2013}. The normalization $Z(t)$ ensures that
\begin{equation}
\Tr(\bar{\rho}(t))=1
\label{normalization}%
\end{equation}
and the thermodynamic conjugates $\{\lambda_j(t)\}$ are chosen in such a way that
\begin{equation}
\Tr(\bar{\rho}(t)A_i) = a_i(t),
\label{macroequivalence}%
\end{equation}
which is called the \ZT{macroequivalence condition}, stating that the relevant density gives the correct mean values for the relevant variables.

Although \cref{rhobar} is a standard choice, a relevant density can be any function of the mean values that satisfies \cref{normalization,macroequivalence} \cite{Grabert1982}. It is helpful, in general, to choose the relevant density in such a way that it is a good approximation for the actual microscopic density, in particular if they coincide for $t=0$.

\subsection{Time-dependent projection operators}
For the time-dependent case, Grabert \cite{Grabert1978,Grabert1982} defines the projection operator by
\begin{equation}
P(t)X = \Tr(\bar{\rho}(t)X) + (A_j - a_j(t))\Tr\!\bigg(\frac{\partial \bar{\rho}(t)}{\partial a_j(t)}X\bigg).
\label{projectiontd}%
\end{equation}
From this form of the projection operator, it is not immediately clear how it is related to the time-independent definition. It can be shown through a longer calculation (see Refs.\ \cite{Grabert1982,teVrugtW2019}) that the operator \eqref{projectiontd} contains the operator \eqref{projection} as a limiting case and that it can be rewritten using a generalized scalar product. However, the general case can be more easily understood from the relevant-density point of view than from the scalar-product point of view. 

We first make clear that $P(t)$ defined by Eq.\ \eqref{projectiontd} is still a projection operator. It has the property
\begin{equation}
P(t)P(t')X = P(t')X.
\label{projectionproperty}%
\end{equation}
This is a generalization of the usual projection operator property $P^2X = PX$. (Note that the property \eqref{projectionproperty} depends on the specific definition \eqref{projectiontd} of the projection operator and can be different for other definitions \cite{Koide2002,teVrugtW2019}.) Moreover, if we continue to think of the relevant variables $\{A_i\}$ as basis vectors in a Hilbert space of operators and add the identity to the set of relevant variables \cite{Grabert1982}, then $P(t)X$, thought of as an element of this Hilbert space, still points in a direction spanned by the relevant variables (including the identity, which gives the first term).

Now, we can use the more general projection operator \eqref{projectiontd} to derive the equations of motion for the relevant variables. For this purpose, we use instead of the Dyson identity \eqref{identity} the more general identity \cite{Grabert1982}
\begin{equation}
\begin{split}
e^{\ii Lt} &=  e^{\ii Lt}P(t) + \INT{0}{t}{s}e^{\ii Ls}(P(s)\ii L Q(s) - \dot{P}(s))G(s,t)\\
&\quad\, +Q(0)G(0,t)
\end{split}\raisetag{2em}%
\label{generaldyson}%
\end{equation}
with the orthogonal dynamics propagator
\begin{equation}
G(s,t)=\exp_R\!\bigg(\ii \INT{s}{t}{t'}LQ(t')\bigg),
\label{odp}%
\end{equation}
where $\exp_R(\cdot)$ denotes a right-time ordered exponential (see \cref{td}). What might be confusing here is that the argument of the exponential is now $LQ(t)$ rather than $QL$ as in the time-independent case. The reason is that, for a time-independent projection operator, we have \cite{teVrugtW2019}
\begin{equation}
\begin{split}
&QG(s,t) = Qe^{\ii LQ(t-s)} \\
&\!=Q\Big(1 + (t-s) \ii LQ + \frac{\ii^2}{2}(t-s)^2LQLQ + \dotsb\Big)\\
&\!=\Big(Q + \ii(t-s)QLQ + \frac{\ii^2}{2}(t-s)^2 QLQLQ + \dotsb\Big)\\
&\!=\Big(1 + \ii(t-s)QL + \frac{\ii^2}{2}(t-s)^2 QLQL + \dotsb\Big)Q\\
&\!=e^{\ii QL(t-s)}Q.
\end{split}
\end{equation}
As we can see, if we compare this to \cref{identity}, we have an additional operator $Q$ at the end. For this reason, the identity \eqref{generaldyson} is applied to $\ii LA$, while the identity \eqref{identity} is applied to $Q \ii L A$.

Applying \cref{generaldyson} to $\ii L A$ gives the equation of motion
\begin{equation}
\begin{split}
\dot{A}_i(t) &= v_i(t) + \Omega_{ij}(t)\delta A_j(t) \\
&\quad\, + \INT{0}{t}{s} \big(K_i(t,s) + \phi_{ij} (t,s)\delta A_j(s)\big) \\
&\quad\, + F_i(t,0),
\end{split}
\label{eqofmotion}%
\end{equation}
where we have introduced the organized drift
\begin{equation}
v_i(t)=\Tr(\bar{\rho}(t)\ii L A_i),
\end{equation}
the fluctuations $\delta A_i(t) = A_i(t) - a_i(t)$, the
collective frequencies
\begin{equation}
\Omega_{ij}(t)=\Tr\!\bigg(\pdif{\bar{\rho}(s)}{a_j(t)}\ii L A_i\bigg),
\end{equation}
the after-effect functions
\begin{equation}
K_i(t,s) = \Tr\!\big(\bar{\rho}(s) \ii LQ(s)G(s,t) \ii L A_i\big),
\label{after}%
\end{equation}
the memory functions
{\begin{align}%
\begin{split}%
\phi_{ij}(t,s) &= \Tr\!\bigg(\frac{\partial \bar{\rho}(s)}{\partial a_j(s)} \ii LQ(s)G(s,t) \ii L A_i\bigg) \\
&\quad\, - \dot{a}_k(s)\Tr\!\bigg(\frac{\partial^2 \bar{\rho}(s)}{\partial a_j(s) \partial a_k(s)} G(s,t)\ii L A_i\bigg),
\end{split}\raisetag{16ex}\label{phiij}%
\end{align}}%
and the random forces
\begin{equation}
F_i(t,0)=Q(0)G(0,t)\ii L A_i.
\end{equation}

In the more general formalism, we get separate equations of motion for the mean values and the fluctuations. From averaging \cref{eqofmotion} we obtain \cite{Grabert1982}\footnote{A more general equation is obtained by considering an initial time $t=u$ rather than $t=0$. The physical significance of $u$ is, in this case, that information about the history of the system from $u$ onwards is taken into account \cite{Grabert1978}.}
\begin{equation}
\dot{a}_i(t) = v_i(t) + \INT{0}{t}{s} K_i(s,t) + f_i(t)
\label{average}%
\end{equation}
with the mean random force $f_{i}(t) = \Tr (\rho(0) F_i(t,0))$. If we assume $\rho(0)=\bar{\rho}(0)$, we have $f_i(t)=0.$ Subtracting \cref{average} from \cref{eqofmotion} gives for the fluctuations \cite{Grabert1982}%
\begin{equation}
\begin{split}
\delta \dot{A}_i(t) &= \Omega_{ij}(t) \delta A_j(t) + \INT{0}{t}{s} \phi_{ij} (t,s) \delta A_j(s) \\
&\quad\:\!+ \delta F_i(t,0)
\end{split}
\label{deltaa}%
\end{equation}
with $\delta F_i(t,0) = F_i(t,0) - f_i(t,0)$. 

Taking a look at \cref{average,deltaa}, we can notice that \cref{deltaa}, which describes the fluctuations, has a structure that is relatively similar to that of \cref{exact}, which we know from the time-independent case. The reason is that, for the fluctuations, we have written down a linear equation. In the close-to-equilibrium case, one can show that mean values and fluctuations follow similar equations of motion, such that \cref{exact} can be recovered by adding them \cite{Grabert1982,teVrugtW2019}. Equation \eqref{average}, on the other hand, is a transport equation for the \textit{mean values} of the observables/operators rather than for the operators themselves, as it was the case in \cref{exact}. For this reason, it is possible that the transport equations are nonlinear. Nonlinear equations do not always make sense for the microscopic observables, which are frequently defined microscopically as sums over delta functions. Moreover, the microscopic observables always follow a linear equation, namely the Liouville equation. No such restrictions hold, in general, for the mean values. 

Of course, obtaining a transport equation for the mean values $\{a_i(t)\}$ is also possible in the time-independent case by averaging over \cref{exact}. This transport equation will then be a linear equation, since the prefactors are time-independent and have no dependence on the macroscopic state determined by the $\{a_i(t)\}$. Although it is always formally possible to apply the time-independent projection operator, it is therefore most useful if one is close to equilibrium, such that thermodynamic nonlinearities are not important \cite{teVrugtW2019}. The case of time-independent projection operators can be recovered from the time-dependent case, if one linearizes the relevant density in the thermodynamic conjugates, i.e., if one assumes deviations from equilibrium to be small \cite{teVrugtW2019,Grabert1982}.

\subsection{\label{td}Time-dependent Hamiltonians}
An even more general case is one in which the Hamiltonian is time-dependent, which, in general, also leads to a time-dependent Liouvillian \cite{teVrugtW2019}. For time-independent projection operators, this case has been considered in Refs.\ \cite{NordholmZ1975,ShibataTH1977,KoideM2000,UchiyamaS1999,Bouchard2007}. Generalizations to time-dependent projection operators are derived in Refs.\ \cite{teVrugtW2019,MeyerVS2019}. We here present the method derived in Ref.\ \cite{teVrugtW2019}, which is applicable also to quantum systems.

This topic requires familiarity with time-ordered exponentials, which students will typically learn about in an advanced quantum mechanics or a quantum field theory course. However, when it comes to the Mori-Zwanzig formalism, certain subtleties become important that are typically ignored in the treatment of time-dependent Hamiltonians. The considerations relevant here thus can be a very valuable ingredient of a quantum mechanics course on time-dependent Hamiltonians, even if one is not interested in their use in statistical mechanics. Here, we explain how this can be done.

The starting point is, again, the Liouville equation
\begin{equation}
\dot{A}(t) = \ii \LH(t)A(t),
\label{liouville}%
\end{equation}
where the Liouvillian is now time-dependent\footnote{See below for a discussion of the subscript H, which denotes the Heisenberg picture.}. For $t > 0$, we can integrate \cref{liouville}, giving
\begin{equation}
A(t) = A_0 + \ii \INT{0}{t}{t'} \LH(t') A(t')
\end{equation}
with $A_0=A(0)$.
This equation can be solved iteratively:
\begin{equation}
\begin{split}
A(t) &= A_0 + \ii \INT{0}{t}{t'} \LH(t') A_0 \\
&\quad\, + \ii^2 \INT{0}{t}{t'}\!\INT{0}{t'}{t''} \LH(t') \LH(t'') A_0 + \dotsb.
\label{second}%
\end{split}
\end{equation}
If we perform this integration infinitely often, we will get an infinite number of terms with increasingly high numbers of Liouvillians. Looking at the second-order expression \eqref{second}, one can already see that, because of the integration boundaries, these are ordered in such a way that the Liouvillian that depends on the latest time is standing on the left. Thus, the Liouvillians are said to be in \textit{left-time order}. Performing infinitely many integrations, we get
\begin{equation}
\begin{split}
A(t) =& A_0 +  \sum_{n=1}^{\infty}\ii^n\INT{0}{t}{t_1}\dotsb\INT{0}{t_{n-1}}{t_n}\LH(t_1)\dotsb\LH(t_n)A_0\\
=&\exp_L\!\bigg(\ii\INT{0}{t}{t'}\LH(t')\bigg)A_0,
\end{split}\raisetag{3em}%
\end{equation}
which defines the \textit{left-time-ordered exponential} $\exp_L(\cdot)$. In analogy, one can also define a \textit{right-time-ordered exponential} $\exp_R(\cdot)$, where later times are standing on the right.

When working with time-dependent Hamiltonians, it is important to distinguish between the \textit{Heisenberg picture} and the \textit{Schr\"odinger picture} of quantum mechanics. In the Schr\"odinger picture, the wave functions or statistical operators are time-dependent, while in the Heisenberg picture, the observables or operators are time-dependent. The operator $A(t)$ in the Heisenberg picture is related to the corresponding Schr\"odinger-picture expression via 
\begin{equation}
A(t) = U^\dagger(t)AU(t)
\label{hs}%
\end{equation}
with the unitary time-evolution operator
\begin{equation}
U(t)=\exp_L\!\bigg(\! -\frac{\ii}{\hbar}\INT{0}{t}{t'}\HS(t')\bigg),
\end{equation}
where $\HS(t)$ denotes the Schr\"odinger-picture Hamiltonian. 
Since the Hamiltonian is time-dependent, the Hamiltonians at different points in time do not necessarily commute, such that there is a difference between the Schr\"odinger-picture Hamiltonian $\HS(t)$ and the Heisenberg-picture Hamiltonian
\begin{equation}
\HH(t) = U^\dagger(t)\HS(t)U(t).  
\end{equation}
As the Liouvillian is defined as the commutator with the Hamiltonian, one therefore needs to distinguish between a Heisenberg-picture Liouvillian $\LH(t)$ and a Schr\"odinger-picture Liouvillian $\LS(t)$, corresponding to the commutators with Schr\"odinger- and Heisenberg-picture Hamiltonians, respectively. Above, we have shown that in terms of Heisenberg-picture Liouvillians, the time evolution can be written as
\begin{equation}
A(t)=\exp_L\!\bigg(\ii\INT{0}{t}{t'}\LH(t')\bigg)A_0.
\label{aexpl}%
\end{equation}
As is shown in Ref.\ \cite{teVrugtW2019}, the time evolution can also be written as
\begin{equation}
A(t)=\exp_R\!\bigg(\ii\INT{0}{t}{t'}\LS(t')\bigg)A_0.
\end{equation}
Remarkably, a right-time-ordered exponential of Schr\"odinger-picture Liouvillians is equivalent to a left-time-ordered exponential of Heisenberg-picture Liouvillians. A direct proof of this is sketched in \cref{appendix}. For the right-time-ordered exponentials, one can prove the identity \cite{teVrugtW2019}
\begin{equation}
\begin{split}
&\exp_R\!\bigg(\ii \INT{0}{t}{t'} \LS(t')\bigg)\\ &=  \exp_R\!\bigg(\ii \INT{0}{t}{t'} \LS(t')\bigg) P(t)\\
&  \quad + \INT{0}{t}{s} \exp_R\!\bigg(\ii \INT{0}{s}{t'} \LS(t')\bigg) \\
& \qquad\qquad\;\:\! \big(P(s) \ii \LS(s)Q(s) - \dot{P}(s)\big) G(s,t)\\
& \quad +Q(0)G(0,t),
\end{split}\raisetag{11ex}%
\end{equation}
which is a generalization of \cref{generaldyson}. Applying this to $\ii \LS(t) A_i$ again gives the general equation of motion \eqref{eqofmotion}.

\section{\label{applications}Applications}
In this section, we present two typical applications of the Mori-Zwanzig formalism to illustrate how it can be used. The Bloch equations are derived using a time-independent projection operator, while dynamical density functional theory is derived with a time-dependent projection operator.

\subsection{Spin relaxation and the Bloch equations}
The treatment of spin relaxation is a standard application of the Mori-Zwanzig formalism \cite{KivelsonO1974,Bouchard2007,teVrugtW2019}. Here, we present a derivation that is a strongly simplified form of the one that can be found in Refs.\ \cite{teVrugtW2019,KivelsonO1974}, which we will follow closely. Details on the spin algebra can be found in Ref. \cite{teVrugtW2019}.

We consider $N$ spins in a time-independent magnetic field $\vec{B} = B_0 \vec{e}_{z}$, where $B_0$ denotes the modulus of the field and $\vec{e}_{z}=(0,0,1)^\mathrm{T}$ its orientation. The system has a total spin $\vec{S}$ given by the sum over the individual spin operators $\{\vec{S}_i\}$. We choose the relevant variables as
\begin{equation}
\vec{A} =
\begin{pmatrix}
S_+\\
\Delta S_z\\
S_- \\
\end{pmatrix}
\end{equation}
with $\Delta S_z = S_z - \braket{S}_{z,\mathrm{eq}}$ (the subscript eq denotes an equilibrium average) and the spin ladder operators $S_\pm = S_x \pm \ii S_y$. Our Hamiltonian reads
\begin{equation}
H =\HL + \HSL - \gamma B_0S_z.
\label{SpinHamiltonian}%
\end{equation}
$\HL$ describes lattice interactions commuting with the spin operator $\vec{S}$, $\HSL$ describes interactions of spin and lattice, and the last term accounts for the interaction with the magnetic field, where $\gamma$ is the gyromagnetic ratio.

If we work in the high-temperature limit, the scalar product of two observables is the expectation value of the product of the two observables. In this case, we can easily calculate the normalization matrix
\begin{equation}
\langle \vec{A} \vec{A}^\dagger \rangle_{\mathrm{eq}} = \frac{1}{4} N \hbar^2
\begin{pmatrix}
2 & 0 & 0\\
0 & 1 & 0\\
0 & 0 & 2\\
\end{pmatrix}
\end{equation}
and the frequency matrix
\begin{equation}
\Omega = - \ii \gamma B_0
\begin{pmatrix}
1 & 0 & 0 \\
0 & 0 & 0\\
0 & 0 & -1\\
\end{pmatrix}.
\label{ommega}%
\end{equation}
For the random force, we find
\begin{equation}
\vec{F}(t) = e^{\ii QLt}\ii Q L \vec{A} = e^{\ii QLt} \frac{\ii}{\hbar} [\HSL, \vec{A}].
\end{equation}
All that is left now is to calculate the memory matrix, the integral over which will -- after a Markovian approximation -- give the dissipative matrix. Due to the typical symmetries of the Hamiltonian, off-diagonal terms of the memory matrix vanish, leading to a decoupling of the equations for the three variables. We obtain, after some calculation, the memory-matrix diagonal elements
{\begin{align}%
\begin{split}%
K_{11}(t) &= \frac{2}{N}\bigg\langle \frac{1}{\hbar^4} \bigg( \! e^{\ii QLt} [\HSL, S_+] \bigg) [S_-,\HSL] \bigg\rangle_{\mathrm{eq}},  
\end{split}\raisetag{7ex}\\
\begin{split}%
K_{22}(t) &= \frac{4}{N}\bigg\langle \frac{1}{\hbar^4} \bigg( \! e^{\ii QLt} [\HSL, S_z] \bigg) [S_z,\HSL] \bigg\rangle_{\mathrm{eq}},  
\end{split}\raisetag{7ex}\\
\begin{split}%
K_{33}(t) &= \frac{2}{N}\bigg\langle \frac{1}{\hbar^4} \bigg(\! e^{\ii QLt} [\HSL, S_-]
\bigg) [S_+,\HSL]
\bigg\rangle_{\mathrm{eq}}. 
\end{split}\raisetag{7ex}%
\end{align}}%
Making the definitions
\begin{align}
\tau_1 =& \INT{0}{\infty}{s}K_{22}(s),\\
\tau_{2,+} =& \INT{0}{\infty}{s}K_{11}(s),\\
\tau_{2,-} =& \INT{0}{\infty}{s} K_{33}(s),
\end{align}
we can, after a Markovian approximation\footnote{To make the presentation simpler, we have ignored one aspect that is discussed in Refs.\ \cite{KivelsonO1974,teVrugtW2019}: For a Markovian approximation to be allowed in the case of spin relaxation, one should first make a transformation to the rotating frame in order to remove fast precession effects, such that one can actually assume the variables to be slow.} and an averaging removing the random force, find the Bloch equations
\begin{align}
\dot{S}_z &= - \frac{S_z - S_{z,\mathrm{eq}}}{\tau_1},\\
\dot{S}_{\pm} &= (\mp \gamma B_0 -  \tau_{2,\pm}^{-1})S_{\pm}.
\end{align}
The terms $\propto \gamma B_0$ describe the precession in a magnetic field. Relaxation towards the equilibrium values is described by the terms $\propto S_z/\tau_1$ and $\propto S_{\pm}/\tau_{2,\pm}$ with the relaxation times $\tau_{1}$ and $\tau_{2,\pm}$.

\subsection{Dynamical density functional theory}
As an example for the application of the time-dependent projection operator technique, we use the derivation of classical dynamical density functional theory (DDFT) \cite{Yoshimori2005,EspanolL2009}. DDFT is a theory for the time evolution of the one-body density $n(\vec{r})$ in a colloidal or atomic fluid which is based on a free-energy functional $F(t)$. While the original derivations have started from Langevin \cite{MarconiT1999,MarconiT2000} or Smoluchowski \cite{ArcherE2004} equations, projection operators have become an important tool in DDFT. In particular, since they can be applied to arbitrary variables, they can (and have been) used to derive extensions of DDFT towards additional variables, such as energy density \cite{WittkowskiLB2012}, entropy density \cite{WittkowskiLB2013}, and momentum density \cite{CamargodlTDZEDBC2018}. 

We suggest DDFT as an example for three reasons. First, it is relatively simple. Second, it is an extremely important theory in soft matter physics, such that students, in addition to learning about projection operators, also learn another method that is of more general importance. Third, DDFT is a good example of a rather general class of nonequilibrium theories known as \ZT{gradient dynamics theories} \cite{Thiele2018,ThieleAP2012}, where the rate of change of a variable or set of variables is proportional to the gradient of the functional derivative of a free energy. Seeing the microscopic derivation of DDFT can thus further contribute to the general understanding of the microscopic origins of irreversible transport equations.

We follow closely the derivation by Espa\~{n}ol and L\"owen presented in Ref.\ \cite{EspanolL2009}.\footnote{A difference in our presentation is that we use, for simplicity, the relevant density \eqref{rhobar}. In Ref.\ \cite{EspanolL2009}, a more general density is used that, in addition, has a factor accounting for the equilibrium configuration.} The considered system consists of $N$ classical particles of mass $m$. As a relevant variable, we choose the number density\footnote{Although $\rho(\vec{r})$ is a more common notation for the number density than $n(\vec{r})$, we here use $n(\vec{r})$ in order to avoid confusion with the probability density.}
\begin{equation}
\hat{n}(\vec{r})=\sum_{i=1}^{N}\delta(\vec{r}-\vec{r}_i),
\end{equation}
where $\vec{r}_i$ is the position of the $i$-th particle. It deserves some comment, since it might not be clear to students, why we are able to derive a field theory for a variable $A(\vec{r},t)$ even though all our considerations were based on variables $\{A_i(t)\}$ that only depend on time. The basic idea is that a field $A(\vec{r},t)$ is an infinite number of variables indexed by the position $\vec{r}$. Thus, we can reuse all previous results for a set of variables $\{A_i\}$. Whenever we encounter a sum over the variables' index, we have to perform an integral over $\vec{r}$, since this corresponds to a sum over all relevant variables. For the same reason, a time derivative $\dot{A}_i$ becomes a \textit{partial} time derivative $\partial A(\vec{r},t)/ \partial t$, since $\vec{r}$ is essentially an index. 

The mean value of $\hat{n}(\vec{r})$ is
\begin{equation}
n(\vec{r},t)=\Tr(\bar{\rho}(t)\hat{n}(\vec{r})),
\end{equation}
where the trace is now an integral over phase space. First, we need to calculate the microscopic current, which in the classical case is done using the Poisson bracket. We find
\begin{equation}
\ii L \hat{n}(\vec{r}) = \sum^{N}_{i=1}(\vec{\nabla}_{\vec{p}_i}H)\cdot(\vec{\nabla}_{\vec{r}_i}\hat{n}(\vec{r})) = - \vec{\nabla}\cdot\vec{J}(\vec{r}) 
\end{equation}
with the current 
\begin{equation}
\vec{J}(\vec{r})=\sum^{N}_{i=1}\frac{\vec{p}_i}{m}\delta(\vec{r}-\vec{r}_i)
\end{equation}
and $\vec{p}_i$ being the momentum of the $i$-th particle. 

Next, we calculate all terms in \cref{average}. We assume $\rho(0) = \bar{\rho}(0)$, which allows to drop the random force $f(t)$. The organized drift is
\begin{equation}
v(\vec{r},t)= -\vec{\nabla}\cdot\Tr(\bar{\rho}(t)\vec{J}(\vec{r}))=0.
\end{equation}
Here, the trace is a phase-space integral that, since the function $\vec{J}$ is odd in the momenta, vanishes.
For the second term, we first use the fact that -- as is shown in Ref.\ \cite{Grabert1982} -- the after-effect function \eqref{after} can be rewritten in the form
\begin{equation}
K_i(t,s)=R_{ij}(t,s)\lambda_j(s),   
\label{relation}%
\end{equation}
where the retardation matrix $R_{ij}(t,s)$ is, for classical systems, given by\footnote{The corresponding quantum-mechanical expression can be found in Refs.\ \cite{Grabert1982,teVrugtW2019}. It is slightly more complicated and not required here. The general structure remains the same.}
\begin{equation}
R_{ij}(t,s)=\Tr(\bar{\rho}(s)(G(s,t)Q(t)\ii L A_i)(Q(s)\ii L A_j)).
\label{retardation}%
\end{equation}
Inserting \cref{relation} into \cref{average} and performing a Markovian approximation gives the equation of motion
\begin{equation}
\dot{a}_i(t)= v_i(t) + D_{ij}(t)\lambda_j(t)
\label{eomMA}%
\end{equation}
with the dissipative matrix\footnote{In the Markovian approximation, one can replace $a_i(s)$ by $a_i(t)$, since variations of $a_i$ are of first order in $\ii L A_i$. The relevant density $\bar{\rho}(t)$ is a functional of the $\{a_i(t)$\}, such that we can also replace $\bar{\rho}(t)$ by $\bar{\rho}(s)$ \cite{Grabert1982}. The same argument applies for the projection operators. Moreover, as in the case of time-independent projection operators, we replace $G(s,t)$ by $\exp(\ii L(t-s))$. Finally, we substitute $s \to t-s$ and switch integration boundaries.}
\begin{equation}
D_{ij}(t)=\INT{0}{\infty}{s}\Tr(\bar{\rho}(t)(Q(t)\ii L A_j)e^{\ii Ls}(Q(t)\ii LA_i)).
\end{equation}
The projected current is
\begin{equation}
Q(t)\ii L\hat{n}(\vec{r})=-\vec{\nabla}\cdot\vec{J}(\vec{r}),
\end{equation}
which gives for the dissipative matrix the expression
\begin{equation}
D(\vec{r},\vec{r}',t)=\vec{\nabla}_{\vec{r}}\cdot(\vec{\nabla}_{\vec{r}'}\cdot M(\vec{r},\vec{r}',t))
\end{equation}
with the mobility
\begin{equation}
M(\vec{r},\vec{r}',t) =\INT{0}{\infty}{s} \Tr(\bar{\rho}(t)\vec{J}(\vec{r}')
\otimes\vec{J}(\vec{r},s)),
\end{equation}
where $\otimes$ denotes a dyadic product. Assuming that the velocities $\{\vec{v}_i=\vec{p}_i/m\}$ of the individual particles are uncorrelated and that the positions vary slowly, this simplifies to\footnote{Since we assume that the velocities vary quickly compared to the positions, we use $\bar{\rho}(t)$ for the expectation value of the positions and $\bar{\rho}(s)$ for the expectation value of the velocities.}
\begin{equation}
\begin{split}
M(\vec{r},\vec{r}',t) &=\INT{0}{\infty}{s} \Tr(\bar{\rho}(s)\vec{J}(\vec{r}')\otimes\vec{J}(\vec{r},s))\\
&\approx\INT{0}{\infty}{s}\sum^{N}_{i,j=1} \Tr(\bar{\rho}(s)\vec{v}_i\otimes\vec{v}_j(s))\\
&\quad\, \Tr(\bar{\rho}(t)\delta(\vec{r}'-\vec{r}_i)\delta(\vec{r}-\vec{r}_j(s)))\\
&\approx\sum^{N}_{i,j=1}D_0\Eins\delta_{ij}\Tr(\bar{\rho}(t)\delta(\vec{r}'-\vec{r}_i)\delta(\vec{r}-\vec{r}_j))\\
&=D_0\Eins n(\vec{r},t)\delta(\vec{r}-\vec{r}')
\end{split}\raisetag{6.6em}%
\end{equation}
with the identity matrix $\Eins$ and the diffusion coefficient
\begin{equation}
D_0 = \frac{1}{3}\INT{0}{\infty}{s}\Tr(\bar{\rho}(s)\vec{v}_i\cdot\vec{v}_j(s)).
\end{equation}
Replacing in Eq.\ \eqref{eomMA} the sum over $j$ by an integral over $\vec{r}'$, this gives
\begin{equation}
\begin{split}
\pdif{n(\vec{r},t)}{t} &= \INT{}{}{^3r'}(\vec{\nabla}_{\vec{r}}\cdot(\vec{\nabla}_{\vec{r}'}\cdot M(\vec{r},\vec{r}',t)))\lambda(\vec{r}',t)\\
&= D_0 \vec{\nabla}_{\vec{r}} \cdot\!\bigg( n(\vec{r},t) \INT{}{}{^3r'}  (\vec{\nabla}_{\vec{r}'} \delta(\vec{r}-\vec{r}')) \lambda(\vec{r}',t)\bigg)\\
&= -D_0\vec{\nabla}\cdot(n(\vec{r},t)\vec{\nabla}\lambda(\vec{r},t)),
\end{split}\raisetag{1.7em}%
\label{n}%
\end{equation}
where in the last step we have integrated by parts.

Finally, we can note that if we define a coarse-grained entropy as
\begin{equation}
S = - k_B\Tr(\bar{\rho}(t)\ln(\bar{\rho}(t))),
\end{equation}
the thermodynamic conjugates $\{\lambda_i\}$ can be written as \cite{Grabert1978}
\begin{equation}
\lambda_i(t)=\frac{1}{k_B}\frac{\partial S}{\partial a_i(t)}.    
\end{equation}
For fields, the partial derivative becomes a functional derivative. Introducing a free energy $F$ by the Legendre transformation
\begin{equation}
F = U - TS    
\end{equation}
with the internal energy $U$ and temperature $T$, we can rewrite \cref{n} as
\begin{equation}
\pdif{n(\vec{r},t)}{t} = \beta D_0 \vec{\nabla}\cdot\!\bigg(n(\vec{r},t)\vec{\nabla}\frac{\delta F}{\delta n(\vec{r},t)}\bigg),
\label{ddft}%
\end{equation}
which is the traditional \textit{DDFT equation}. 

Since $D_0 >0$, we can easily prove the H-theorem
\begin{equation}
\begin{split}
\tdif{F}{t} &= \INT{}{}{^3r}\frac{\delta F}{\delta n(\vec{r},t)}\pdif{n(\vec{r},t)}{t}\\
&=\INT{}{}{^3r}\frac{\delta F}{\delta n(\vec{r},t)}\beta D_0 \vec{\nabla}\cdot\!\bigg(n(\vec{r},t)\vec{\nabla}\frac{\delta F}{\delta n(\vec{r},t)}\bigg)\\
&= -\INT{}{}{^3r}\beta D_0 n(\vec{r},t)\bigg(\vec{\nabla}\frac{\delta F}{\delta n(\vec{r},t)}\bigg)^2 \leq 0.
\end{split}
\end{equation}
We have thus -- by virtue of restricting ourselves to one relevant variable and the approximation that this variable shows Markovian behavior -- obtained an irreversible dissipative law, the DDFT equation, from the reversible Hamiltonian dynamics we started with. Moreover, the Mori-Zwanzig formalism provides a microscopic expression for the free energy. 

The structure of \cref{ddft}, where the dynamics of the relevant variable is driven by the gradient of a thermodynamic conjugate, is a linear-response equation and a very general property of close-to-equilibrium systems. This can be shown with the Mori-Zwanzig formalism in a more general way \cite{WittkowskiLB2013}. General treatments also allow to show how generic features of nonequilibrium thermodynamics, such as irreversibility or the Onsager relations, arise from the microscopic physics in the Mori-Zwanzig formalism \cite{Grabert1982}.

\section{\label{conclusion}Conclusions}
In this article, we have provided a compact introduction to the Mori-Zwanzig formalism, including both the standard, time-independent projection operator formalism and more advanced forms including time-dependent projection operators and time-dependent Hamiltonians. Particular attention has been paid on points that are, for someone who is new to the field, potentially difficult to understand, such as the relation between the scalar product and the relevant density approach. Relevant applications from the literature to spin relaxation and DDFT have also been presented.

We believe that the Mori-Zwanzig formalism, presented in this way, can form a valuable ingredient of basic and advanced courses in statistical mechanics and quantum mechanics. It is, in its simple forms, relatively easy to explain, and allows to introduce many important general ideas of theoretical physics, such as the origin and general structure of dissipative transport equations in statistical mechanics and the relation between Schr\"odinger picture and Heisenberg picture.

\section*{Acknowledgements}
R.W.\ is funded by the Deutsche Forschungsgemeinschaft (DFG, German Research Foundation) -- WI 4170/3-1.

\appendix
\section{\label{dyson}Proof of the Dyson identity}
We here present the standard derivation of the Dyson identity \eqref{identity}, which can be found, e.g., in Ref.\ \cite{Bouchard2007}. Consider the quantity
\begin{equation}
W(t) = e^{-\ii Lt}e^{\ii QLt}
\label{w}%
\end{equation}
and take the time derivative:
\begin{equation}
\dot{W}(t)= - e^{-\ii Lt}P\ii L e^{\ii QLt}.
\end{equation}
Integrating this with respect to time from 0 to $t$ gives
\begin{equation}
W(t) = W(0) - \INT{0}{t}{s}e^{-\ii Ls}P\ii L e^{\ii QLs}.
\label{w(t)}%
\end{equation}
We use the initial condition $W(0)=1$, insert \cref{w} into \cref{w(t)}, and multiply from the left by $e^{\ii Lt}$. This gives
\begin{equation}
e^{\ii QLt} = e^{\ii Lt} - \INT{0}{t}{s}e^{\ii L(t-s)}P\ii L e^{\ii QLs},
\end{equation}
which is the Dyson identity.

\section{\label{appendix}Relation between Heisenberg picture and Schr\"odinger picture}
The identity \cite{teVrugtW2019}
\begin{equation}
\exp_R\!\bigg(\INT{0}{t}{t'}\ii\LS(t')\bigg)=\exp_L\!\bigg(\INT{0}{t}{t'}\ii\LH(t')\bigg),  
\label{transformation}%
\end{equation}
based on which a Mori-Zwanzig formalism for time-dependent Hamiltonians can be derived, can be proven in two ways. An indirect proof based on expectation values that is valid to all orders is given in Ref.\ \cite{teVrugtW2019}. Here, we sketch a different proof that is not written down there. It is more lengthy, but conceptually easier, since it is based on a direct calculation.

We show the derivation up to second order, which is the first nontrivial one. The Heisenberg-picture Hamiltonian reads
\begin{equation}
\begin{split}%
\HH(t) &= U^\dagger(t)\HS(t)U(t)\\
&= \exp_R\!\bigg(\frac{\ii}{\hbar}\INT{0}{t}{t'}\HS(t')\bigg)\HS(t)\\
&\quad\,\exp_L\!\bigg(\! -\frac{\ii}{\hbar}\INT{0}{t}{t'}\HS(t')\bigg).
\end{split}%
\end{equation}
Up to second order, we can thus write
\begin{equation}
\begin{split}%
\HH(t) &= \HS(t) + \frac{\ii}{\hbar}\INT{0}{t}{t'}\HS(t')\HS(t)\\
&\quad\, -\frac{\ii}{\hbar}\INT{0}{t}{t'}\HS(t)\HS(t') + \dotsb
\end{split}%
\end{equation}
Using Eq.\ \eqref{aexpl}, this gives for an arbitrary operator $A$ the relation 
{\allowdisplaybreaks\begin{equation}
\begin{split}%
&\exp_L\!\bigg(\INT{0}{t}{t'}\ii\LH(t')\bigg)A\\
&=\bigg(1+\INT{0}{t}{t'}\ii\LH(t')\\
&\quad\, + \INT{0}{t}{t'}\!\INT{0}{t'}{t''}\ii\LH(t')\ii\LH(t'') + \dotsb \bigg)A\\
&= A + \frac{\ii}{\hbar}\INT{0}{t}{t'}[\HH(t'),A]\\
&\quad\, +\Big(\frac{\ii}{\hbar}\Big)^2\INT{0}{t}{t'}\!\INT{0}{t'}{t''}[\HH(t'),[\HH(t''),A]] + \dotsb \\
&= A + \frac{\ii}{\hbar}\INT{0}{t}{t'}[\HS(t'),A]\\
&\quad\,+\Big(\frac{\ii}{\hbar}\Big)^2\INT{0}{t}{t'}\!\INT{0}{t'}{t''}[\HS(t'')\HS(t'),A]\\
&\quad\,-\Big(\frac{\ii}{\hbar}\Big)^2\INT{0}{t}{t'}\!\INT{0}{t'}{t''}[\HS(t')\HS(t''),A]\\
&\quad\,+\Big(\frac{\ii}{\hbar}\Big)^2\INT{0}{t}{t'}\!\INT{0}{t'}{t''}[\HS(t'),[\HS(t''),A]] + \dotsb .
\end{split}\raisetag{12em}%
\end{equation}}
The terms of second order in $\HS$ can be rewritten as
{\allowdisplaybreaks\begin{equation}
\begin{split}%
&[\HS(t'')\HS(t'),A] - [\HS(t')\HS(t''),A] \\
&\quad\,+ [\HS(t'),[\HS(t''),A]] \\
&=\HS(t'')\HS(t')A
+A\HS(t')\HS(t'') \\
&\quad\, -\HS(t')A\HS(t'')-\HS(t'')A\HS(t') \\
&=[\HS(t''),[\HS(t'),A]],   
\end{split}%
\end{equation}}
such that
{\allowdisplaybreaks\begin{equation}
\begin{split}%
&\exp_L\!\bigg(\INT{0}{t}{t'}\ii\LH(t')\bigg)A\\ 
&=A + \frac{\ii}{\hbar}\INT{0}{t}{t'}[\HS(t'),A]\\
&\quad\, +\Big(\frac{\ii}{\hbar}\Big)^2\INT{0}{t}{t'}\!\INT{0}{t'}{t''}[\HS(t''),[\HS(t'),A]] + \dotsb \\
&=\bigg(1+\INT{0}{t}{t'}\ii\LS(t')\\
&\quad\, +\INT{0}{t}{t'}\!\INT{0}{t'}{t''}\ii\LS(t'')\ii\LS(t') + \dotsb \bigg)A\\
&=\exp_R\!\bigg(\INT{0}{t}{t'}\ii\LS(t')\bigg)A.
\end{split}\raisetag{9em}%
\end{equation}}
Since the operator $A$ is arbitrary, we obtain from this Eq.\ \eqref{transformation}. 
We have thus shown, up to second order, that replacing the Heisenberg Liouvillian by the Schr\"odinger Liouvillian corresponds to switching from left time order to right time order.

\nocite{apsrev41Control}\clearpage
\bibliographystyle{apsrev4-1}
\bibliography{control,refs}
\end{document}